\documentclass[11pt,a4paper]{article}

\pdfoutput=1

\usepackage{jheppub}
\usepackage{amsmath}
\usepackage{amsfonts}
\usepackage{graphicx}
\usepackage{amssymb}
\usepackage{amssymb}
\usepackage{latexsym}
\usepackage{array}
\usepackage{bbold}
\usepackage[normalem]{ulem}
\usepackage{xcolor}
\usepackage{slashed}
\usepackage{cancel}

\newcommand{\real}{{\mathop{\mathrm{Re}}}}

\newcommand{\bea}{\begin{eqnarray}}
\newcommand{\eea}{\end{eqnarray}}
\newcommand{\be}{\begin{equation}}
\newcommand{\ee}{\end{equation}}
\newcommand{\re}[1]{(\ref{#1})}

\usepackage{xcolor}
\definecolor{mygreen}{HTML}{006E28}

\title{Quasi Q-Balls in massless scalar fields}

\author[a]{Tomasz Roma\'nczukiewicz}
\author[b]{and Yakov Shnir}
\affiliation[a]{Faculty of Theoretical Physics, Astronomy and Applied Computer Science, Jagiellonian University, Krak\'ow, Poland}
\affiliation[b]{Instituto de F\'{i}sica de S\~{a}o Carlos; IFSC/USP;
Universidade de  S\~{a}o Paulo, USP\\
Caixa Postal 369, CEP 13560-970, S\~{a}o Carlos-SP, Brazil}

\emailAdd{tomasz.romanczukiewicz@uj.edu.pl}
\emailAdd{shnir@maths.tcd.ie}

\abstract{We study long-term evolution of radiating quasi-Q-balls in 1+1 dimensional models without mass threshold. Two different models are considered, 
the model with a rational modification of the usual Q-ball sextic potential and the model of a Q-ball in a box with outgoing boundary conditions. 
We find that the outgoing boundary conditions modify the angular frequency of quasi-Q-balls which becomes complex-valued. The quasi-Q-balls decay initially very similarly, but in the last stage quasi-Q-balls in a box become linear resonances, whereas in the rationally modified model, they decay according to some power law. 
}

\begin{document}

\maketitle

\section{Introduction}

Recent years have seen great progress in constructing and analysis of spatially
localized regular solutions of nonlinear field equations, the solitons,
for a review, see e.g. \cite{manton2004topological,shnir2018topological}.
A particular class of solitons, known as Q-balls,
represents time-dependent configurations of a self-interacting complex scalar field
with a stationary oscillating phase \cite{Rosen:1968mfz,Friedberg:1976me,Coleman:1985ki,Radu:2008pp}.
Q-balls are non-topological solitons, they carry a conserved
Noether charge which is related to an unbroken, continuous global symmetry of the corresponding Lagrangian \footnote{A vast literature is devoted to
the  self-gravitating stationary spinning solitons, the boson stars, which we will not discuss here.}
(for recent reviews, see \cite{Radu:2008pp,Nugaev:2019vru}).

Typical examples of Q-balls are spherically symmetric solutions of the renormalizable
Friedberg-Lee-Sirlin two-component model with a symmetry breaking potential \cite{Friedberg:1976me}, or solutions
of the non-linear model with a single complex
scalar field and a sextic self-interaction potential \cite{Coleman:1985ki}. Physically,
such configurations can be considered as a condensate of a large number of scalar quanta which, for a
fixed value of the Noether charge, corresponds to the stationary points of the effective energy functional. In such
a context, the charge, which is proportional to the angular frequency,
can also be interpreted as the particle number.

Notably, Q-balls exist  within a certain
angular frequency range, which is determined by the
explicit structure of the potential. Typically,
there are two branches of  solutions, which are
represented by two curves of the dependencies of the
energy of the configuration on its charge \cite{Friedberg:1976me}. Q-balls are
stable against particle emission along the lower branch, as their mass is smaller than
the perturbative mass of free quanta of scalar excitations \footnote{However,
quantum corrections may lead to instability of small Q-balls \cite{Graham:2001hr}.}.

The problem of stability of the Q-balls has been extensively studied in the literature, see e.g.
\cite{Tsumagari:2008bv,Graham:2001hr,Sakai:2007ft,Tranberg:2013cka,Mai:2012yc}. It was shown that for
some range of values of the parameters of the model there
are stable, metastable and unstable solutions. Three different stability criteria are usually
considered \cite{Tsumagari:2008bv}:
(i) Stability against fission, which is similar to a generic Vakhitov-Kolokolov criterion;
(ii) Stability with respect to  decay into free scalar quanta, which happens as the angular frequency increases above
upper critical value; (iii) Classical
stability with respect to  perturbations of the fields. In particular,
the last case covers perturbations of the Q-ball via incoming radiation \cite{Saffin:2022tub,Ciurla:2024ksm}.
However, corresponding analysis is quite involved since the spectrum of excitations
may include both normal and quasinormal (QNM) modes \cite{Ciurla:2024ksm}, moreover, in
some cases the excitations even cannot be considered as linearized perturbations \cite{Smolyakov:2017axd}.

There is a certain similarity between the long-time evolution of stationary Q-balls and
oscillons \cite{Bogolyubsky:1976nx,Copeland:1995fq,
Gleiser:1993pt}, which are very long-lived, almost periodic non-linear field configurations. Roughly speaking,
Q-ball can be viewed as a system of two coupled oscillons, associated with the real
components of a complex scalar field \cite{Copeland:2014qra}. This analogy is supported by the
existence of an approximately conserved adiabatic invariant \cite{Kasuya:2002zs,Kawasaki:2015vga,Levkov:2022egq}.
Notably, slowly radiating oscillons pass through a sequence of quasi-stable Q-ball-like configurations
\cite{Honda:2001xg,Dorey:2019uap,Zhang:2020bec}. The reverse is also true, an oscillon can be considered as a pair of Q-ball and anti Q-ball \cite{blaschke2025qballpolarizationsmooth}.
Some, most recenct development include establishing the link between oscillons and Q-balls via renormalization group \cite{blaschke2024oscillonsqballsrenormalization}.

Studying of Q-balls becomes simpler in 1+1 dimensional models \cite{Axenides:1999hs,Battye:2000qj,Bowcock:2008dn}, there is an exact
analytical solution of the stationary field equations. In a recent paper \cite{Ciurla:2024ksm} we considered perturbative excitations
of such solitons, in particular, we analyzed the spectral structure of the excited Q-balls. It has been observed that the spectrum
contains both scattering and localized modes, as well as the half-propagating modes, related to the so-called
quasinormal modes.

The aim of this paper is to investigate the existence of Q-ball-like structures in models without mass threshold. 
We will refer to  such configurations as \textit{quasi-Q-balls} (QQBs) because they inherit  qualitative features from both Q-balls and quasi-normal modes.

We show that the QQBs arise in a certain class of theories with a rational modification of the potential, which supports the usual Q-balls. We also examine a somewhat simplified version of the theory, where the potential is modified in space in  such a way, that inside a
restricted spacial domain, a box,  the equations of motion are exactly the same as in some standard Q-ball model. However, outside the box the field satisfies linear massless wave equation, so the configuration becomes unstable due to outgoing radiation.

\section{Models}
\subsection{Standard Q-balls}
We consider a complex scalar field theory in $(1+1)$ dimensions
described by the following Lagrangian
\cite{Axenides:1999hs,Battye:2000qj,Bowcock,Ciurla:2024ksm}
\begin{equation}
    \mathcal{L} = \partial_\mu\phi\partial^\mu\phi^*-V\left(|\phi|^2\right)\, ,
    \label{lag}
\end{equation}
where the asterisk denotes complex conjugation and the rescaled potential of the self-interacting complex scalar field is
\begin{equation}
    V\left(|\phi|^2\right)=|\phi|^2-|\phi|^4+\beta |\phi|^6\,.
    \label{pot}
\end{equation}
The model has one free parameter $\beta$, as $\beta > 1/4$, there is a single global minimum at $\phi=0$, for $\beta < 1/3$ there is a second local minimum at
\begin{equation}
    |\phi_\textrm{min}|^2=\frac{1+\sqrt{1-3 \beta}}{3\beta}
\end{equation}
and a maximum at 
\begin{equation}
    |\phi_\textrm{max}|^2=\frac{1-\sqrt{1-3 \beta}}{3\beta}\,.
\end{equation}

If $\beta < 1/4$ the potential becomes unbounded from below and the local miminum at $\phi=0$ turns out to be  a false vacuum. In the critical limit $\beta=1/4$ the vacuum is twofold degenerate, $V(|\phi|)=0$ as $|\phi|=\{0,\sqrt 2 \}$ and the model \re{lag} also supports topological solitons, the kinks.  

This simple model can serve as a prototype example for study of Q-balls. Indeed, the Lagrangian \re{lag} is invariant with respect to global $U(1)$ transformations $\phi \to \phi e^{i\alpha}$, the corresponding Noether current is  
\begin{equation}
   \label{curr}
j_\mu = i(\phi^* \partial_\mu \phi -\partial_\mu \phi^*\, \phi ) \, .
\end{equation}

The field equation of the model \re{lag} 
\be
\phi_{tt}-\phi_{xx}+\frac{\partial V}{\partial |\phi|^2}\phi=0
\label{field_eq}
\ee
can be solved using the usual Q-ball parametrization 
$\phi(x, t) = e^{i\omega t}f(x)$, where $\omega$ is the internal rotation frequency and $f(x)$ is a real profile function which satisfies the first order equation   
\begin{equation}
    \label{stationary}
    \frac{df}{dx}=\pm\sqrt{\tilde V(f^2)}
\end{equation}
with the boundary conditions $f(x\to\pm \infty) =0$. 
Here the effective potential is given by
\begin{equation} \label{effective_pot}
    \tilde V(f^2) = V(f^2) -\omega^2 f^2 \, .
\end{equation}
The exact solution of Eq. \re{stationary}
is \cite{Axenides:1999hs,Battye:2000qj,Bowcock}
\begin{equation}
    f(x;\omega) = \frac{\sqrt{2}\omega'}{\sqrt{1+\sqrt{1-4\beta\omega'^2}\cosh(2\omega'x)}}\, ,
\label{solution}
\end{equation}
where $\omega'=\sqrt{1-\omega^2}$ is the complementary frequency \cite{Bowcock}.
Hence, the total energy and the charge of the stationary configuration are  
\begin{equation}
    E = \frac{4\omega {\omega'} + Q (4\beta-1 + 4 \beta \omega^2 )}{8\omega \beta}\,  ;\qquad
Q = \frac{4\omega}{\sqrt \beta}{\rm arctanh}\left(\frac{1-\sqrt{1-4\beta {\omega'}^2}}{2{\omega'} \sqrt \beta}\right) \, .
\label{EQ}
\end{equation}

Localized solutions of the model \re{lag} exist for some finite range of the angular frequency, $\omega_\textrm{min}\le \omega \le \omega_\textrm{max}$, where the upper bound in 1+1d corresponds to the mass threshold,   
$\omega_\textrm{max}=V^{\prime}(0)=1$. The lower bound depends on the parameter $\beta$ \cite{Bowcock}, 
\begin{equation}
    \omega_{\rm min}(\beta) = \sqrt{1-\frac{1}{4\beta}}\, .
    \label{wmin}
\end{equation}

\begin{figure}
    \centering
    \includegraphics[width=0.85\textwidth]{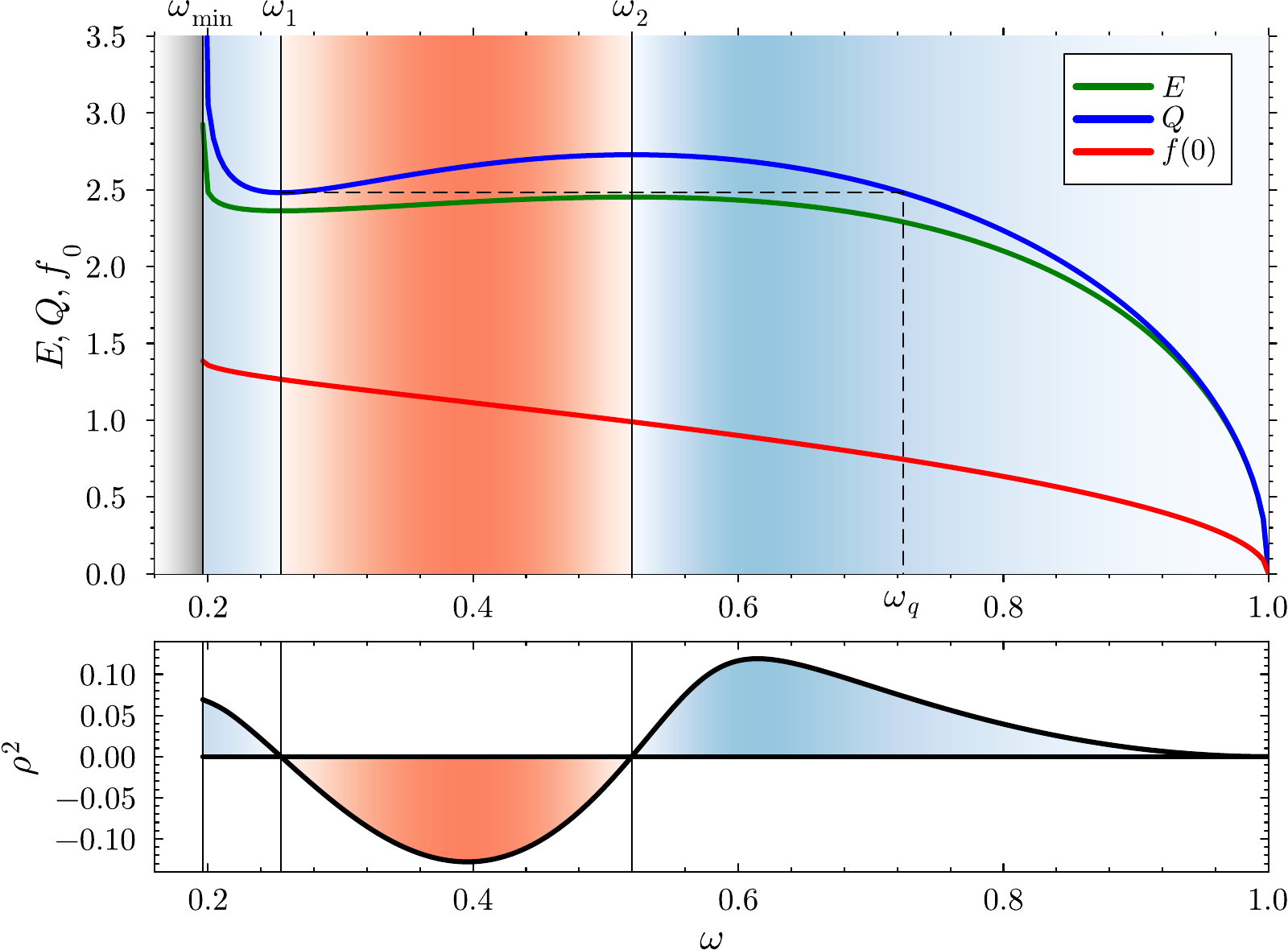}    
    \caption{Q-ball stability in the $\phi^6$ model for $\beta=0.26$. Blue shaded regions indicate stable Q-balls, red shaded unstable. }\label{fig:Qball_stability}
\end{figure}

As a particular example, in Fig~\ref{fig:Qball_stability} we displayed the curves of the dependencies of the energy, the charge and value of the Q-ball profile function $f$ at the centre of the configuration, as functions of angular frequency $\omega$ for $\beta=0.26$. The Q-ball is classically stable with respect to small perturbations
if \cite{Anderson70, Friedberg:1976me,Lee:1991ax,Tsumagari:2008bv}
\be\label{stability}
\frac{\omega}{Q}\frac{dQ}{d\omega}<0 \, .
\ee
This relation, as one can see from Fig~\ref{fig:Qball_stability}, holds for the two blue shaded domains of values of angular frequency $\omega$. The red shaded domain corresponds to unstable Q-balls. 

The instability of the Q-ball becomes manifest as we consider perturbations of the form 
\be
\phi(x,t) = f(x)e^{i\omega t} + \delta \phi(x,t) 
\ee
where \cite{Ciurla:2024ksm}
\be
\delta \phi(x,t) = \eta_1(x) e^{i(\omega+\rho)t}+\eta_2(x) e^{i(\omega-\rho)t}
\ee
Here, the complex parameter $\rho$ corresponds to the perturbations of the basic frequency of the Q-ball. If the condition \ref{stability} is satisfied, the Q-balls are stable and in the spectrum of small perturbations there is no unstable mode. If the stability condition is not fulfilled, Q-balls poses a single unstable mode with purely imaginary perturbation eigenvalue $\rho$ (or $\rho^2<0$), excitation of which leads to an exponential growth of perturbation (see  Fig~\ref{fig:Qball_stability}, bottom plot). 

\subsection{Massless model}
\label{massless}
Let us  consider a rational modification of the potential of the 
Q-ball model above. In analogy with a similar discussion of the time evolution of oscillons \cite{Dorey:2023sjh}, we  
choose the deformed potential 
\begin{equation}
 V\left(|\phi|^2;\epsilon\right)=\frac{W^2}{\epsilon+W}
 \label{deform}
\end{equation}
where
\begin{equation}
W\left(|\phi|^2\right)  =|\phi|^2-|\phi|^4+\beta |\phi|^6\,.
\label{potphi6}
\end{equation}
Note that the equation of motion takes the following form
\begin{equation}
    \phi_{tt}-\phi_{xx}+\frac{dV}{d|\phi|^2}\phi=0\,,
\end{equation}
therefore for configurations with $|\phi|$ constant in space
\begin{equation}
m_{\textrm{eff}} = \sqrt{\frac{dV}{d|\phi|^2}}   
\end{equation}
can play the role of an effective mass.

\begin{figure}
    \centering
    \includegraphics[width=1\textwidth]{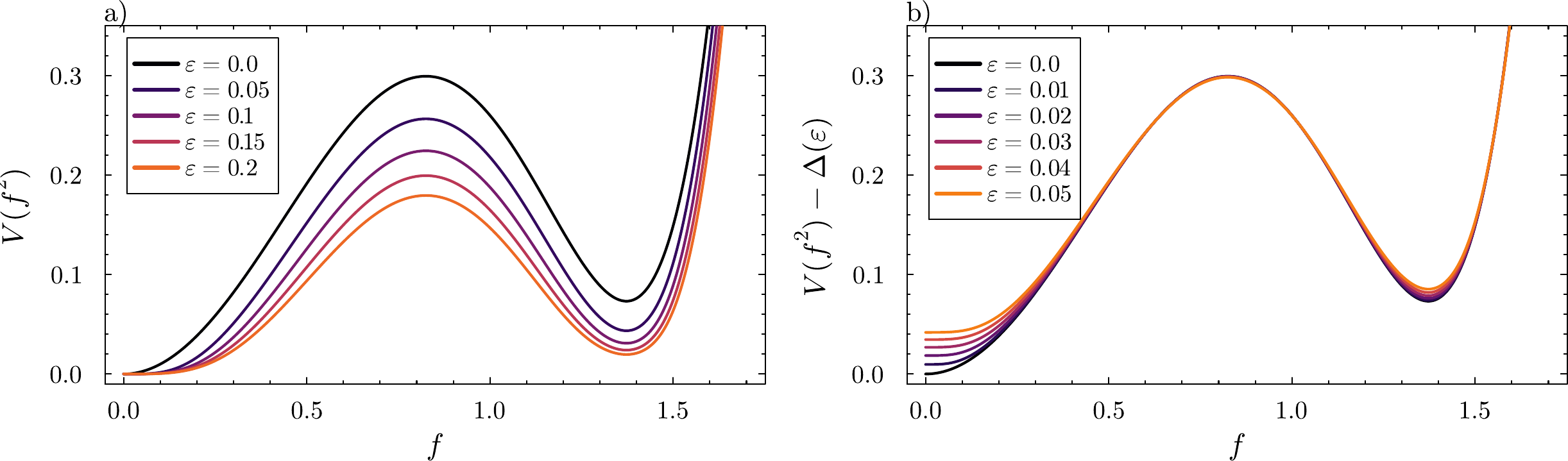}
    \caption{Potentials for different values of $\epsilon$ (a) and potentials shifted by $\Delta(\epsilon)=V(f^2_\textrm{max};0)-V(f^2_\textrm{max};\epsilon)$ (b)}
    \label{fig:massless_potentials}
\end{figure}

The modification changes the effective mass of the small perturbation around the vacuum so that $m_{\textrm{eff}}=0$, compare with $m_{\textrm{eff}}=1$ for $W$. The potential \re{potphi6} depends on two parameters, $\beta$ and $\epsilon$. The latter parameter $\epsilon$ is responsible for near the vacuum deformation of the original model \re{lag}.  The potentials are plotted in Figure \ref{fig:massless_potentials} a).
The smaller $\epsilon$ the smaller range of $|\phi|>0$ where the model is significantly different form $W(|\phi|^2)$. In order to emphasize the similarities we plotted the potentials $V(|\phi|^2)$ shifted by $\Delta(\epsilon)=V(f^2_\textrm{max};0)-V(f^2_\textrm{max};\epsilon)$ so that the maxima would coincide (Figure \ref{fig:massless_potentials} b)).
However, any non-zero value of $\epsilon$ yields flatness of the potential \re{potphi6} in the vicinity of the vacuum. Consequently,  the configuration becomes unstable, there is no mass gap and some amount of energy leaks into radiation continuously. 
The left plot in Fig.~\ref{fig:rational_vs_box} displays the snapshot of the profile function of the quasi-Q-ball at $\epsilon=0.005,~ \beta=0.26$ and $\omega=0.6$, the oscillating radiative tail and the core of the configuration are clearly visible.  

The analogy between oscillons and Q-balls in 1+1 dimensions suggests that the lifetime of the massless Q-ball can be large. The configuration consists from a central core and radiative tails in the far field zone, the perturbative parameter $\epsilon$ controls the magnitude of the radiation which is vanishing in the limit $\epsilon=0$.

\subsection{Q-ball in a box}

\begin{figure}
    \centering
    \includegraphics[width=1\textwidth]{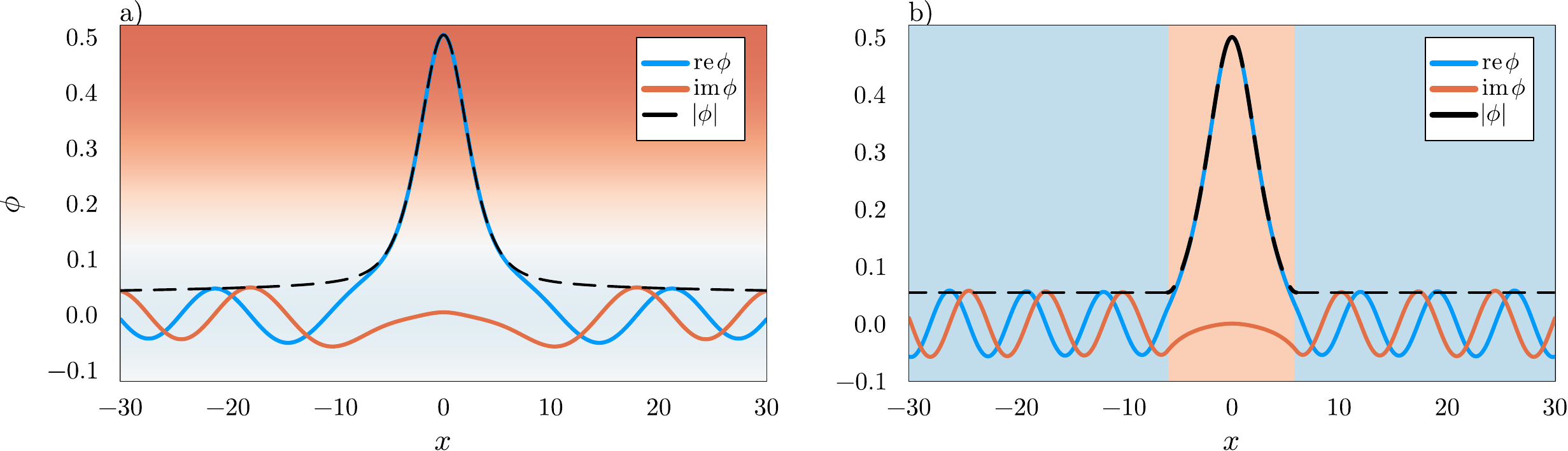}
    \caption{ Comparison of the profile functions of the quasi-Q-ball in the deformed massless model \re{deform} at $\epsilon=0.005,~ \beta=0.26$ (a) and the quasi-Q-ball in the box with outgoing boundary conditions at $L=6$ (b) for $\phi(0, 0)=0.5$.}
    \label{fig:rational_vs_box}
\end{figure}

In order to analyse time evolutions of a radiating quasi-Q-ball, it will be convenient to 
confine the configuration 
into 
a finite region of space, \textit{the box},  bordered by two open segments outside. Inside the box $x\in[-L,L]$
the field obeys the usual equations of motion supporting ordinary Q-balls (\ref{field_eq}) while in the outer region, the dynamics of a scalar field is governed by the massless wave equation  $\phi_{tt}-\phi_{xx}=0$. Such a setup does not allow stationary Q-balls to exist because of the radiation flux escaping from the box. However, if the box is large enough, the radiation losses can be small, so the object inside the box can live sufficiently long. 

In Fig.~\ref{fig:rational_vs_box}b), we displayed the snapshot of the profile function of the "boxed" Q-ball at $L=6$ and $\omega=0.6$.  

Let us first focus on a small amplitude approximation. 
Naively, the field equation inside the box  can be approximated with its linearized version, the usual Klein-Gordon equation with mass parameter $m=1$: 
\begin{equation}
    \phi_{tt}-\phi_{xx}+\phi=0\quad \textrm{for}\quad |x|<L.
\end{equation}
The separation of variables gives
\begin{equation}\label{qqb_profile}
    \phi(x,t) = e^{i\omega t}\psi(x)
\end{equation}
where, for a parity-even solution, satisfying purely outgoing boundary conditions
\begin{equation}
    \psi(x) = 
    \begin{cases}
        A\cos(kx)& \textrm{for} \quad x\leq|L| \\
        Be^{-i\omega |x-L|}& \textrm{for} \quad x\geq|L|
    \end{cases}
    \label{outBC}
\end{equation}
$k=\sqrt{\omega^2-1}$ is a wave number inside the box. Note, that this time, $\psi(x)$ is complex valued, therefore we used a different symbol to distinguish from the real valued profile $f(x)$ of the ordinary Q-Balls.
By matching the values and derivatives on the boundary of the regions we obtain a condition for $\omega$
\begin{equation}
    k\tan(kL)=i\omega
\end{equation}
Clearly, this transcendental equation can be solved numerically. For a particular value $L=6.5$ (which we will mostly use in our consideration below), 
the first four solutions are characterized by the frequencies gathered in Table \ref{tab:resonances}.
\begin{table} \centering
    \begin{tabular}{ccc}               
        \hline\hline
        \hspace*{1cm}$n$\hspace*{1cm}&\hspace*{1cm}$\Omega$\hspace*{1cm}&\hspace*{1cm}$\Gamma$\hspace*{1cm}\\ 
        \hline
        1 & \texttt{1.0269} &  \texttt{0.0083} \\ 
2 & \texttt{1.2249} &  \texttt{0.0595} \\ 
3 & \texttt{1.5534} &  \texttt{0.1199} \\ 
4 & \texttt{1.9490} &  \texttt{0.1711} \\ 
        \hline
    \end{tabular}
    \caption{The first few quasinormal modes of quasi Q-ball $\omega=\Omega+i\Gamma$ for $L=6.5$.}\label{tab:resonances}
\end{table}
The frequencies are complex-valued therefore the solutions are quasinormal modes (or resonances). We will often split the frequency in real and imaginary part $\omega=\Omega+i\Gamma$.
We expect that the lowest resonance is the final stage of the decay of Q-balls.
We will denote the quasinormal mode frequencies $\omega_{\rm res}$.

Note that, for $|x|>L$ 
\begin{equation}
    \psi(x)=Be^{-i\Omega|x-L|}e^{\Gamma|x-L|}
\label{psi}
\end{equation}
which means that the solutions corresponding to complex frequencies possess exponentially growing tails as $|x|\to\infty$ and cannot be normalized. This is a well-known feature of resonances in quantum mechanics and particle physics \cite{lamb1900peculiarity, Landau:1991wop}. Such resonances are just a very useful mathematical idealization of physical states, which must be described by normalizable wave functions, which locally are very close to the profiles of the QNMs. Complex eigenvalues are signs of broken hermicity of the operator describing linear perturbation due to the outgoing wave boundary conditions, which violate the time reflection symmetry. Most common applications include a description of radioactivity and relaxation of black holes, i.e. the process with a clearly defined arrow of time.

\subsection{Quasi-Q-balls as nonlinear resonances} 

Let us assume now that the approximation (\ref{qqb_profile}) is also valid for large amplitude. However, in the nonlinear case we expect that the profiles would depend on frequency, $\omega$ as is the case for ordinary Q-balls. In the case of a Q-ball in a box or in the massless case we cannot assume that the solutions would be stationary but rather, because of the radiation, the objects would decay changing their shapes and frequencies. Therefore, we need to assume that even in the approximate case, the frequency is time dependent. 
Plugging the approximation describing
\begin{equation}
    \Phi(x,t;\omega(t)) = e^{i\omega(t) t}\psi(x;\omega(t))
\end{equation}
into the equation of motion we obtain
\begin{equation}    
    -\omega^2\Phi-\Phi_{xx}+\frac{\partial V}{\partial|\Phi^2|}\Phi+
    2i\dot\omega\left(\Phi+\omega\frac{\partial\Phi}{\partial\omega}\right)+
    \ddot\omega\frac{\partial\Phi}{\partial\omega} + \dot\omega^2\frac{\partial^2\Phi}{\partial\omega^2}=0
\end{equation}
When $\omega(t)$ is a slowly varying function so that $\ddot{\omega}\ll\dot\omega\ll1$ we can assume that, in the first approximation, the Q-balls satisfy the standard equation of motion
\begin{equation}    
    -\omega^2\Phi-\Phi_{xx}+\frac{\partial V}{\partial|\Phi^2|}\Phi\approx0 \label{QQBEquation}
\end{equation}

In the same fashion as in the linearized case we can approximate the solution with the form
\begin{equation}\label{QQBsolution}
    \Phi(x,t)=e^{i\omega t}\psi(x)
\end{equation}
and impose outgoing boundary conditions for $|x|=L$ also in the nonlinear case, assuming that the field in the bulk has the form of a monochromatic wave, 
\begin{equation}
    \psi'(L)=-i\omega \psi(L)\label{match}
\end{equation}
allowing the frequency to be complex along with all the consequences mentioned above. Due to exponentially growing tails, QQBs have infinite energy and charge when considered on an infinite domain. However, the energy and the charge are finite when considered inside the box $|x|<L$. 

Note that the ansatz  (\ref{qqb_profile}) is just an adiabatic approximation since a QQB slowly radiates the energy away and its frequency changes as it decays. This approximation is justified because the decay rate is small and the QQB inside the box is almost periodic.
Clearly, the matching condition \re{match} 
together with outgoing boundary condition \re{outBC}, 
implies that the outgoing wave in the far field zone  $|x|\gg L$ still carries all the information about earlier dynamics of the QQB inside the box.

\begin{figure}
    \centering
    \includegraphics[width=1\textwidth]{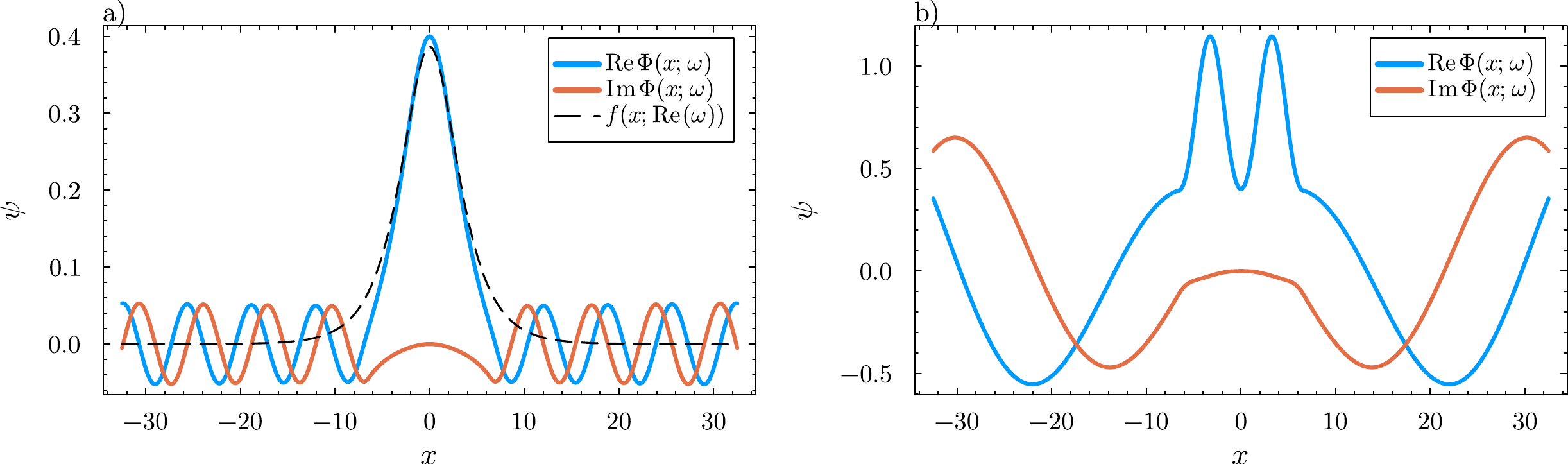}
    \caption{The  real and imaginary parts of the solution to (\ref{QQBEquation}) are shown for the illustrative solutions  at $\omega=0.9255 + 0.0031i$ (a) and $\omega=0.1919 + 0.0198i$ (b).}
    \label{fig:examplePQballs}
\end{figure}

The outgoing boundary conditions allow for more than one branch of boxed QQBs. Typical profiles of two such branches are shown in Figure \ref{fig:examplePQballs} 
with initial condition $\psi(0)=f_0=0.5$. We will focus on the QQBs with the lowest imaginary part of $\omega$, which have the longest lifespan and have a single and prominent maximum at $x=0$. 

Note that the solution displayed in Figure \ref{fig:examplePQballs} b) looks like two Q-balls stuck closely together. In general, Q-balls which are in phase repel and such a configuration is highly unstable. However, the radiation escaping from the box exerts pressure on the Q-balls pushing them towards each other and making the configuration more stable.

\begin{figure}
    \centering
    \includegraphics[width=0.75\textwidth]{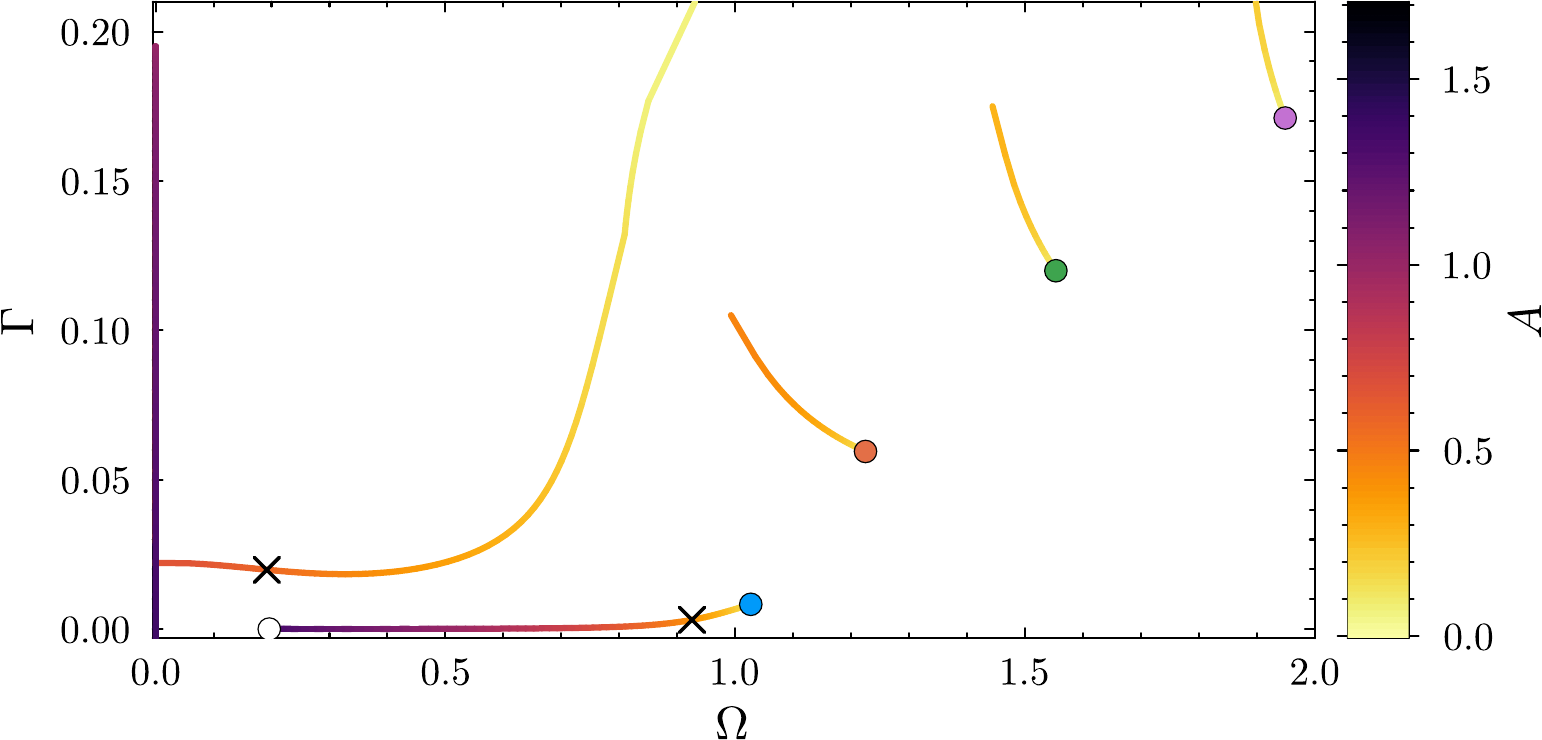}
    \caption{Disconnected branches of solutions for $L=6.5$. Two crosses correspond to the solutions displayed in Fig.\ref{fig:examplePQballs}, the colored circles indicate the starting solutions listed in Table \ref{tab:resonances}.}
    \label{fig:examplePQballsBranches}
\end{figure}

In Figure~\ref{fig:examplePQballsBranches} we display the disconnected branches of the decaying solutions which originate from the four quasinormal modes listed in Table \ref{tab:resonances}. Note that the lower branch exists for all allowed ranges of values of the angular frequency.

As seen in Fig~\ref{fig:examplePQballsBranches}, there are also resonance solutions displayed on the vertical axis for $\Omega=0$ and  $\Gamma>0$. In a linearized theory, these solutions correspond to so-called anti-bound states. We are not certain if they may play any significant role in the full dynamics.

\begin{figure}
    \centering
    \includegraphics[width=1\textwidth]{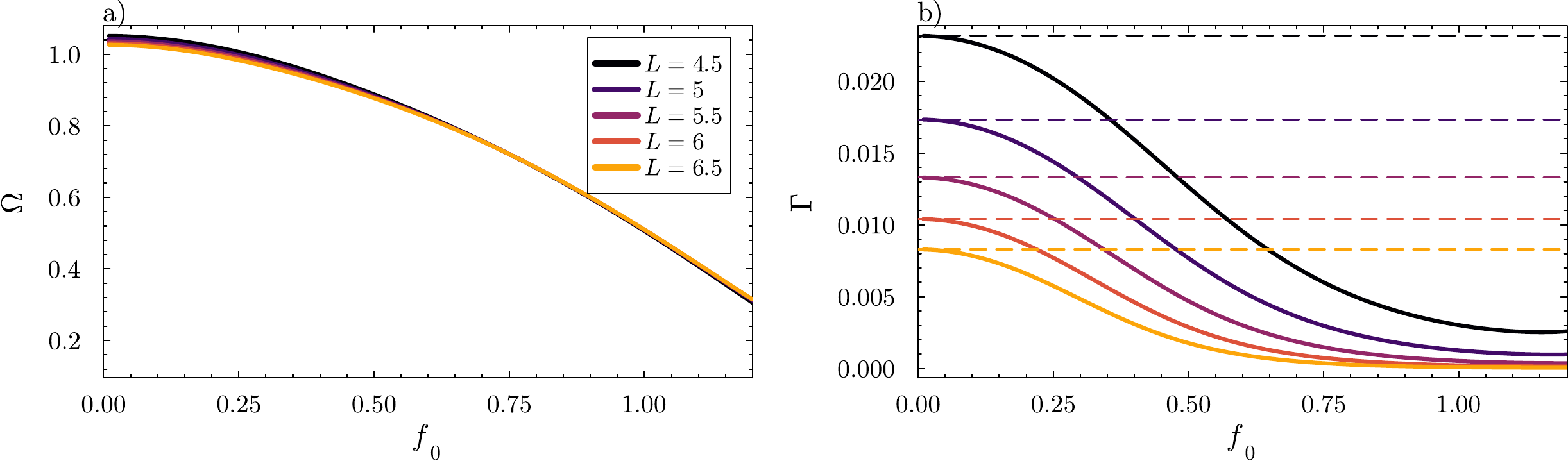}
    \caption{Example frequencies (real part (a) imaginary part (b)) of the QQBs found for different values of the size of the box $L$ as a function of the value of the field at the centre $f_0$ for the most stable branch and $\beta=0.26$. Dashed lines correspond to the lowest QNMs. }
    \label{fig:examplePQballsfreqs}
\end{figure}

\begin{figure}
    \centering
    \includegraphics[width=0.5\textwidth]{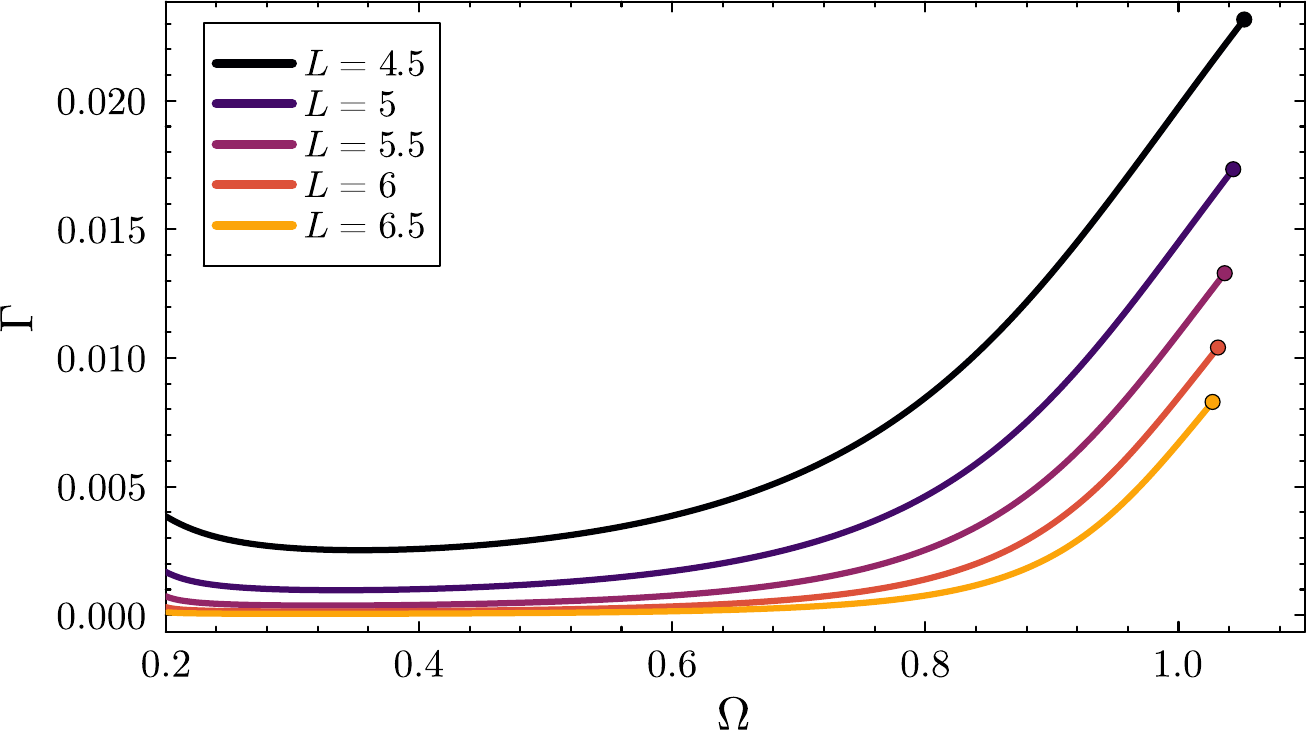}
    \caption{Example frequencies real part and imaginary part (b) of the QQBs found for different values of the size of the box $L$ for the most stable branch and $\beta=0.26$. Points correspond to the lowest QNMs. }
    \label{fig:examplePQballsfreqsb}
\end{figure}
In the case of complex frequencies the usual formulae for the charge and  the energy of the configuration are modified as 
\begin{equation}
    \label{Q-charge}
Q= \int dx\, j_0 = (\omega+\omega^*) \int dx\, \psi\psi^* = 2\Omega \int dx\, |\psi|^2\ ,
\end{equation}
\begin{equation}\label{Energy}
    E=\int dx  \left\{ |\psi_x|^2 + |\omega|^2|\psi|^2 + V(|\psi|^2) \right\}
\end{equation}
However, for $L=6.5$ both the energy and the charge inside the box do not differ significantly from the original theory, and the corresponding counterpart of the Figure \ref{fig:Qball_stability} looks almost the same up to the width of the lines. 
On the other hand, the imaginary part of the frequency, $\Gamma$ highly depends on the box size and on the amplitude of the QQB. As shown in Figure \ref{fig:examplePQballsfreqs}b) for large amplitude QQBs (with $\phi(0,0)=f_0\approx 1$) have small $\Gamma$ which grows as the amplitude decreases. In the limit $f_0\to0$, $\Gamma$ tends to a maximal value which corresponds to the
 imaginary part of the lowest quasi-normal mode frequency.

\subsection{Decay rate of boxed Quasi-Q-balls}

\begin{figure}
    \centering
    \includegraphics[width=1\textwidth]{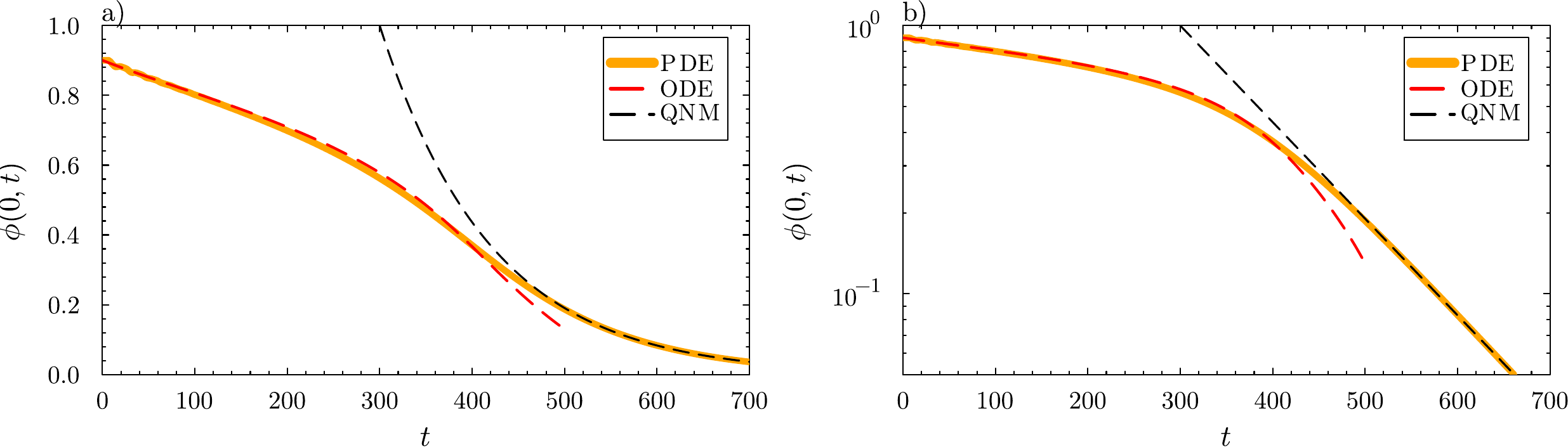}
    \caption{Comparison between the numerical simulations of the full model (orange), solution to (\ref{ODE_E}) red dashed line, linear QNM decay (dashed black) and decay from equation (\ref{ODE_Gamma}). Linear scale (a) and logarithmic scale (b). $L=0.6$ and initial conditions for PDE correspond to a standard QB with $\omega=0.6$.}
    \label{fig:compareODE}
\end{figure}

First, we consider radiating quasi-Q-balls in a box.
One may expect that the imaginary part of $\omega$ gives the  instantaneous rate of the decay of the QQBs
\begin{equation}\label{ODE_Gamma}
   \phi(x,t)=e^{-\Gamma t}e^{i\Omega t}\psi(x)\quad \Rightarrow\quad\frac{df_0}{dt}=-\Gamma(f_0)f_0.
\end{equation}
However, this is not supported by numerical simulations in general, see Fig.\ref{fig:compareODE}. Such an approximation is correct only in the small amplitude limit. The reason is that the large amplitude Q-balls may have different shapes and a simple scaling factor does not transform one QQB into another with a different size. 

A better approximation to the time evolution of large QQBs can be obtained from the condition of conservation of energy. 
The energy is radiated away from the boundary of the box, 
so that the energy which stays in a box changes as 
\begin{equation}\label{ODE_E}
    \frac{dE}{dt} = -\mathcal{P}(L)+\mathcal{P}(-L)=-4|\omega\psi(L)|^2\,.
\end{equation}
The rate of change of the frequency can be calculated from the following ordinary differential equation
\begin{equation}\label{freq_evolution}
    \frac{d\Omega}{dt} = \frac{1}{dE/d\Omega}\frac{dE}{dt} 
\end{equation}
Since the frequency $\Omega$ of the QQBs weakly depends on the size of the box, see Fig.~\ref{fig:examplePQballsfreqs}, we can 
approximate the value of the field inside the box using the ansatz \re{qqb_profile} with $\omega \approx \Omega$ for sufficiently large box sizes. 
However, this approximation becomes less accurate as 
$\Omega$ tends to $\real\,\omega_{\rm res}>1$.

Since ordinary Q-balls exist only for $\omega<1$ equation \re{freq_evolution} breaks at $\Omega=1$. But for $\Omega>1$ the evolution of the QQBs is well approximated with the decay of the QNM. All three scenarios, i.e. approximations  
(\ref{ODE_E}), \re{freq_evolution} and the linear QNM decay, compared to the direct numerical simulations, are shown in Figure \ref{fig:compareODE}. Since the QQBs possess exponentially growing tails with infinite energy, we chose stationary solutions corresponding to standard QB  as initial conditions for the full PDE and observed the relaxation process. 

\begin{figure}
    \centering
    \includegraphics[width=1\textwidth]{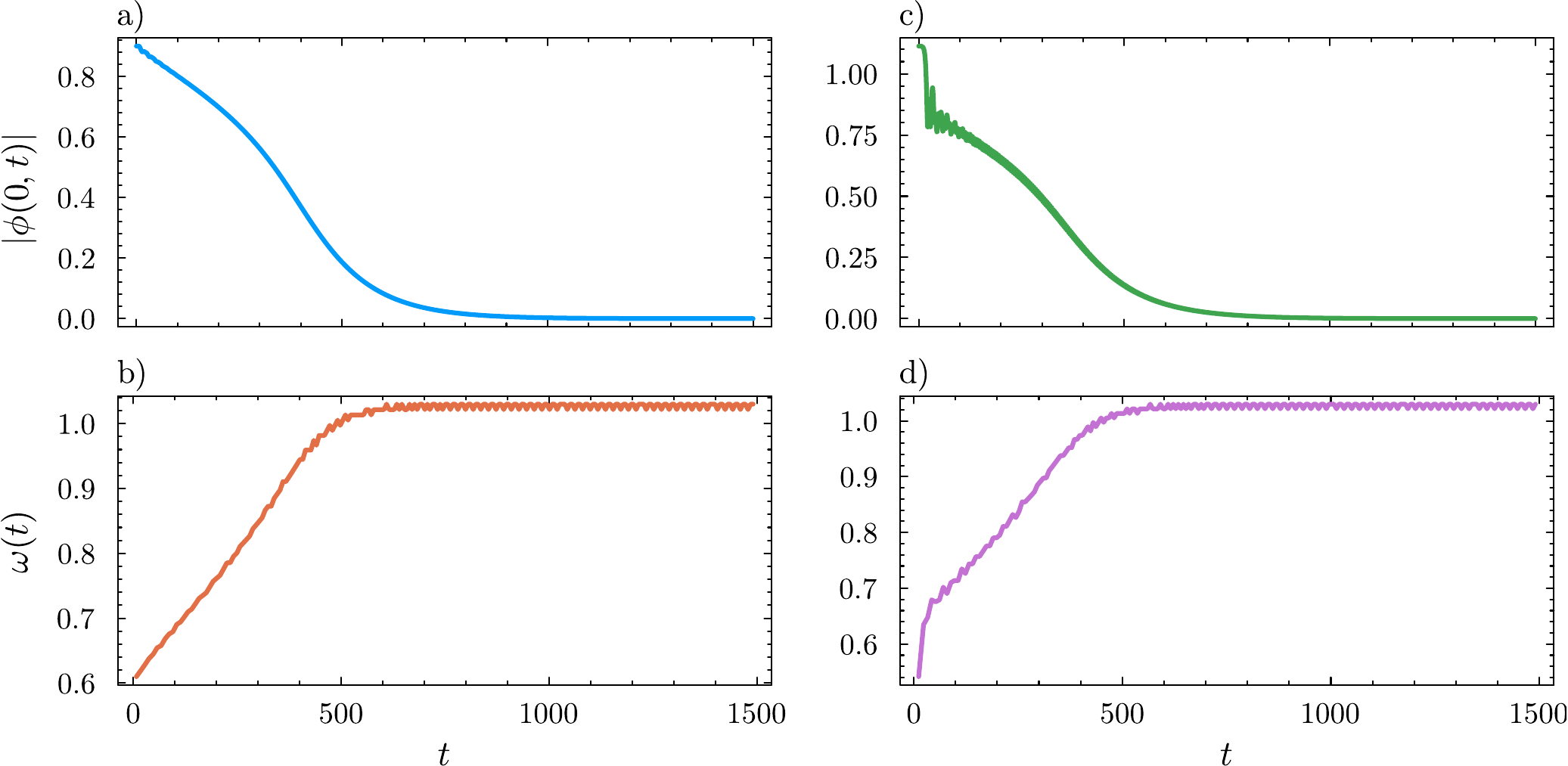}
    \caption{Boxed QQB decay for $L=6.5$ and $\omega=0.6$ (a,b) and  $\omega=0.4$ (c,d).}\label{fig:decay06_domain}
\end{figure}

\begin{figure}
    \centering
    \includegraphics[width=0.75\textwidth]{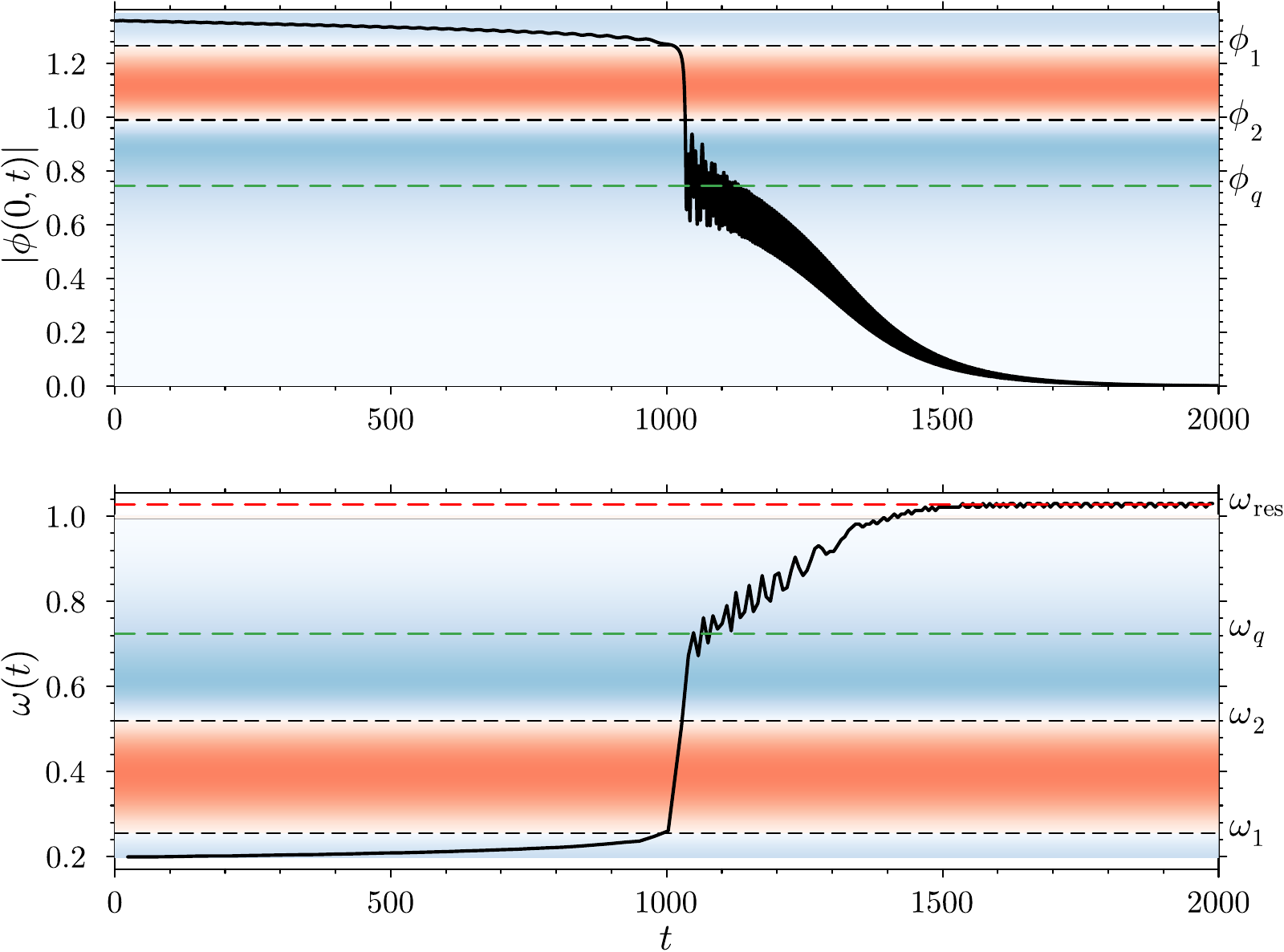}
    \caption{Q-ball decay in a box for $\omega=0.2$, $L=6.5$ -- different stages.}\label{fig:decay02_domain}
\end{figure}

Considering the long time evolution of QQBs, as a particular example we take the parameter 
$\beta=0.26$, because, in such a case, 
there are three different stability intervals of $\omega$, see Fig~\ref{fig:Qball_stability}.
The minimal allowed value of the frequency is $\omega_{\rm min}=0.196116$ and  $\frac{dQ}{d\omega}$ has two zeros for $\omega_1=0.255100$ and $\omega_2=0.519694$.
Note, that $dE/d\omega=0$ for $\omega=\omega_{1,2}$.
Therefore, we have to consider three subintervals:
\begin{itemize}
 \item $\omega\in(\omega_2, 1)$. In this case the Q-balls in the original model \re{lag},\re{pot} are stable but QQBs in a box radiate some amount of energy away and slowly decay
 increasing their frequency until it reaches the frequency $\omega_{\textrm{res}}$ (see Figure \ref{fig:decay06_domain} a,b).

 \item $\omega\in(\omega_1, \omega_2)$. Then the Q-balls in the original model \re{lag},\re{pot}  are unstable. This instability is dominating the first, rather short phase of the decay, during which a Q-ball decays into a stable Q-ball with a similar charge, the frequency $\omega_q=0.724441$ and smaller amplitude and energy. $Q(\omega_q)=Q(\omega_1)$. This phase is indicated by a dashed line in Fig~\ref{fig:Qball_stability}. The energy is transformed into bound, quasinormal and usual scattering modes. After this phase, the decay is similar to the one observed in the stable case described above. The only difference is an additional oscillations due to the excitation of the eigenmodes. (see Figure \ref{fig:decay06_domain} c,d).

 \item  $\omega\in(\omega_{\rm min},   \omega_1)$. In such a case the Q-balls in the original model \re{lag},\re{pot}  are stable. The decay of the QQBs in a box has three distinctive phases, see Figure \ref{fig:decay02_domain}. First, the quasi-Q-ball decays increasing its frequency. But this phase lasts only until the Q-ball reaches the critical frequency $\omega_1$ and the field at the centre reaches $\phi_1=f(\omega_1)$. Then the configuration becomes unstable and, repeating the above-mentioned scenario, after a relatively short transition towards the solution on the second stable branch with a frequency $\omega_q=0.724441$, the configuration continues to decay towards the mass threshold, see Figure \ref{fig:decay02_domain}.
 
\end{itemize}

\begin{figure}
    \centering
    \includegraphics[width=0.75\textwidth]{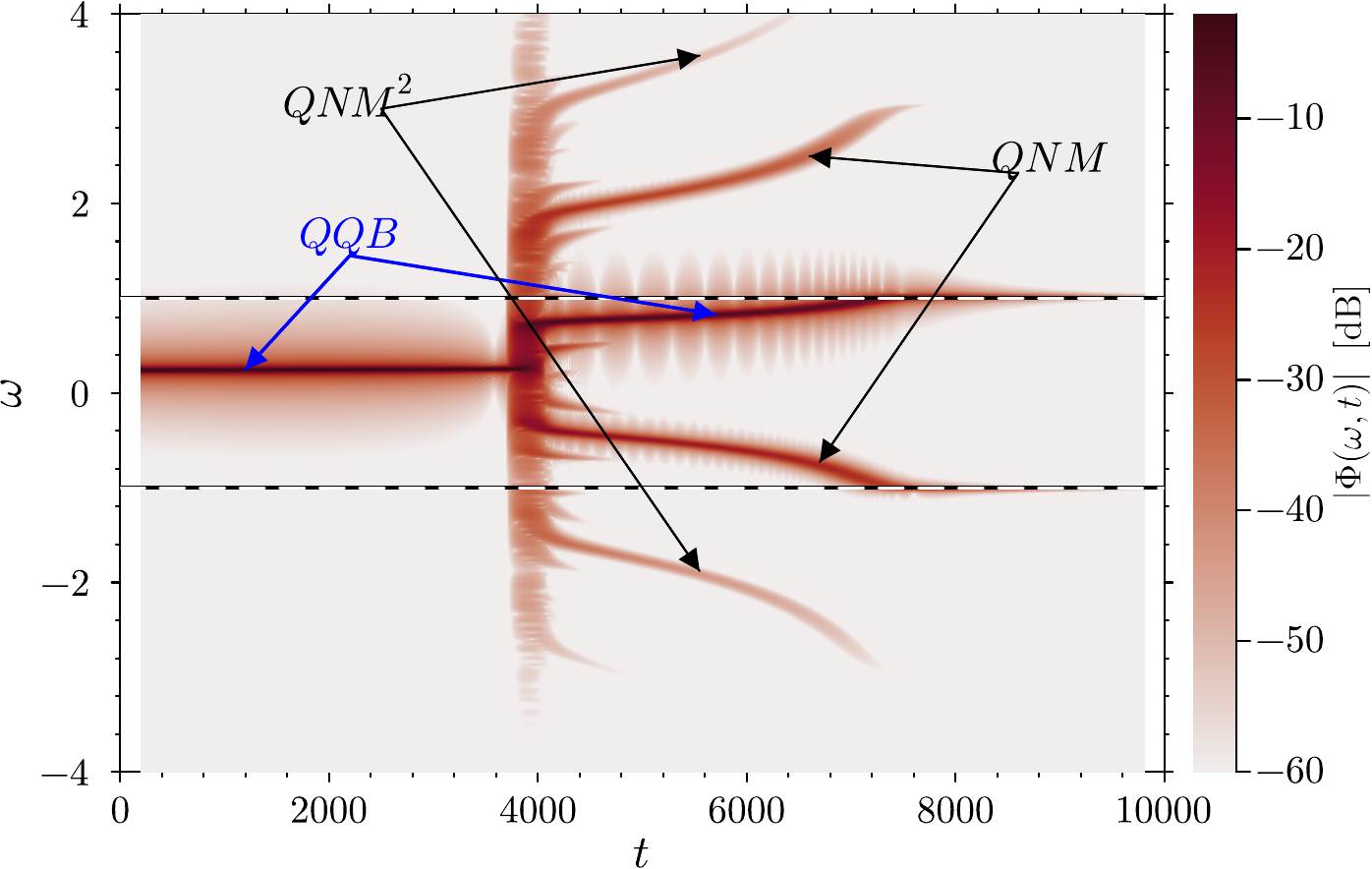}
    \caption{Time-frequency spectrogram for initial $\omega=0.24$ and $L=9$ showing the evolution of Q-ball in a box with visible peaks for Q-ball's fundamental frequency, $QQB$, quasi-normal mode, $QNM$ and its second harmonic, $QNM^2$. Dashed white lines indicate the mass threshold.}\label{fig:spectrogram_box}
\end{figure}

In Figures \ref{fig:decay06_domain}-\ref{fig:decay02_domain} we have shown the decays of QQB measuring the amplitude of the field at the origin and the dominant frequency by measuring positions of subsequent maxima of the real part of the field. Such analysis gives information about the dominant frequency of the system but does not give the full picture. 
Figure \ref{fig:spectrogram_box} shows the time-frequency domain spectrogram of the field value at the centre for a relatively large box $L=9$ ensuring slow evolution allowing good resolution both in frequency and time domains. Initial conditions correspond to a standard Q-ball with $\omega=0.24$, close to the end of the first stable branch. For about  4000 unites of of time the QQB evolved slowly, radiating out its energy and charge. This stage is visible as a dark, single line on spectrogram \footnote{The apparent oscillations of the width of the line is a simple numerical artifact of the discrete FFT. If the frequency matches exactly the frequency sampling of the FFT it becomes a very narrow peak, but when the frequency is in between discrete frequencies of the FFT the peak is spread because such non-matching frequency can be represented only as a superposition of many discrete oscillations.}. At around $t=4000$, when the QQB reached the end of the stable branch, a sudden transition to the second stable branch is visible. After the transition, the fundamental frequency was approximately $\omega_q$. In the spectrogram other dark lines appear, indicating that the QQB after the transition was highly excited. Some of the excitations are radiated out and vanish relatively quickly. However, except for the fundamental frequency, four dark lines are clearly visible for a substantial amount of time. Two of them situated symmetrically around the fundamental frequency of the QQB can be identified as a quasinormal mode. In \cite{Ciurla:2024ksm} such modes were described in more detail for standard QB. These QNMs, characterized by a single complex eigenvalue, $\rho$, have two components oscillating with $\omega\pm\rho$. Imaginary parts are surprisingly small (of order $10^{-5}$), which means that such modes can live very long. The real part of $\omega-\rho$ is within the mass gap. The profile of the component corresponding to this frequency is localized. Real part of $\omega+\rho$ is outside the mass gap, and the corresponding component is responsible for the decay of the QNM. In that sense, such QNMs are Feschbach resonances. Both components change their frequencies as the QQB evolves adiabatically. They are labeled $QNM$ on the spectrogram. There are two other dark lines, visible of the spectrogram, positions of which can be identified as $\omega\pm2\rho$. Such frequencies can appear in the dynamical system due to quadratic nonlinearities, therefore they are labelled as $QNM^2$.  Note that the QNM vanish as the QQB decays and becomes a QNM itself.

\subsection{Evolution in the massless model}
Finally, we shortly discuss the time evolutions of the radiating Q-balls in the massless\footnote{We remark that the  terminology "massless" 
refers to the linearized scalar excitations in the asymptotically far region. However, by analogy with the slowly radiating oscillons \cite{Mukaida:2016hwd,Levkov:2022egq,Dorey:2023sjh} the effective mass of excitations in the interior of the large Q-ball is different.}
 model  \re{lag},\re{deform}. The mechanism of the energy loss due to radiation is somewhat
different from what is outlined above for the Q-balls in a box.  In the limit $\epsilon \to 0$ the deformed potential \re{deform} is not quadratic at the vacuum  \cite{Dorey:2023sjh} and, as mentioned above in the section \ref{massless}, the mass of linearized excitations is vanishing. Consequently, for some small non-vanishing value of the parameter $\epsilon$,  there is a continuous leakage of the energy of the Q-ball to radiation.

\begin{figure}
    \centering
    \includegraphics[width=1\textwidth]{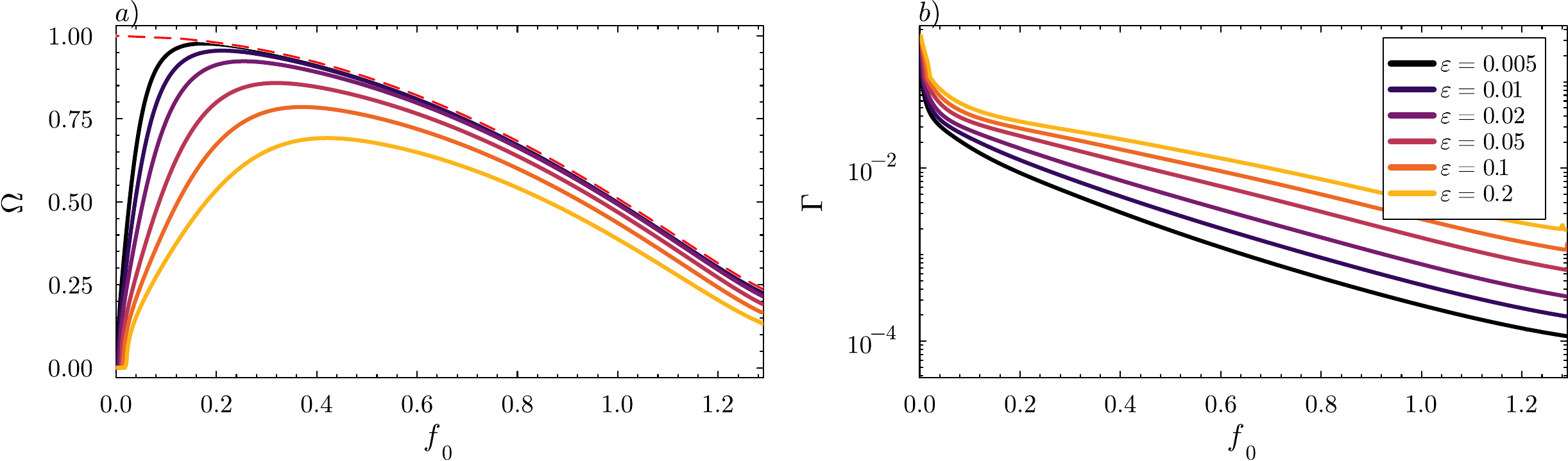}
    \caption{Frequency (a) and the damping term dependence on the amplitude of the QQB in a massless model. The dashed red line corresponds to the standard QB ($\epsilon=0$).}\label{fig:massless_freqs}
\end{figure}

The solutions satisfying the outgoing boundary condition for nonlinear resonance also give somewhat different results (Figure \ref{fig:massless_freqs}) than those for QQBs in a box (Figure \ref{fig:examplePQballsfreqs}). Firstly, there is no box, therefore there does not exist any linear QNM as the small amplitude limit. Linearization around vacuum simply gives a free wave equation. Instead, as the amplitude decreases, so does the frequency $\Omega$, but the damping term $\Gamma$ increases; see Figure \ref{fig:massless_freqs}. Large amplitude QQBs are similar to the standard QB and large QQB in a box i.e. as the amplitude becomes larger, the frequency $\Omega$ decreases. As a result, for every value of $\epsilon$ there is a maximal frequency of the QQB. Numerical calculations show that this maximal frequency is below the mass threshold of the unmodified mode ($m=1$ for $\epsilon=0$). 

\begin{figure}
    \centering
    \includegraphics[width=0.5\textwidth]{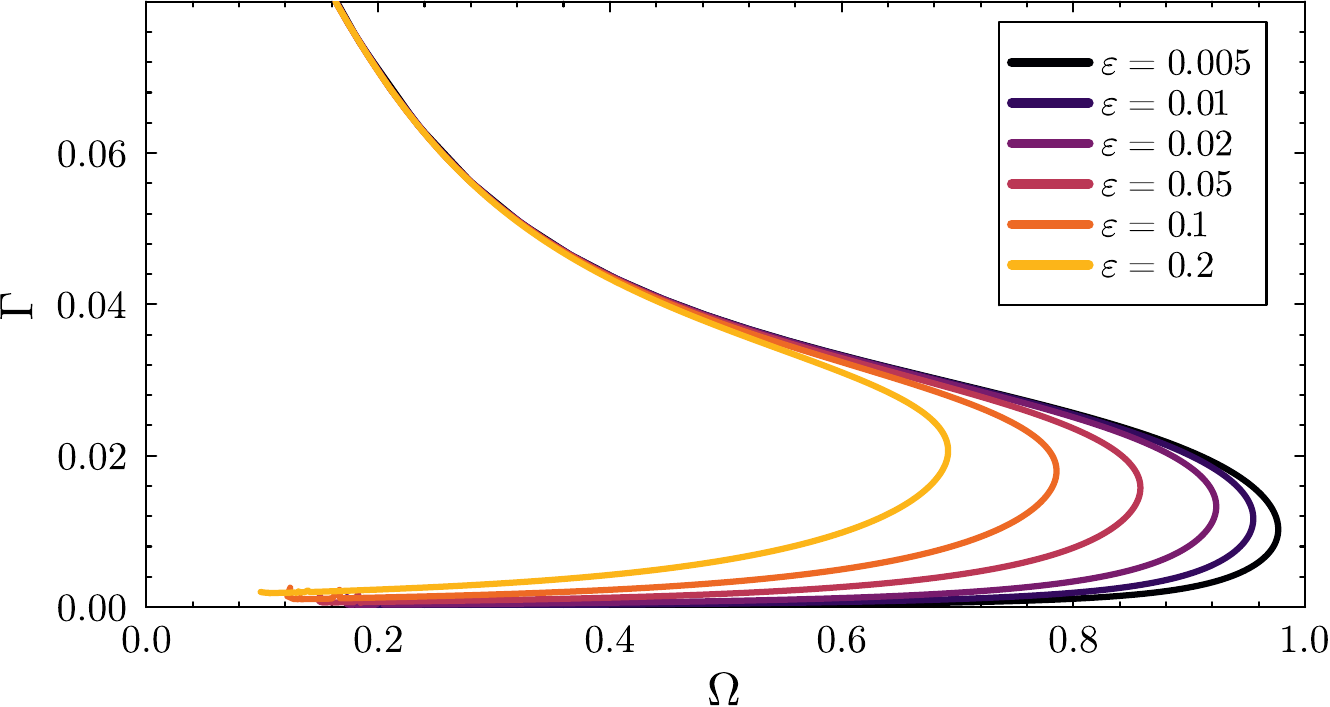}
    \caption{Example frequencies real part and imaginary part (b) of the QQBs found for different
values of $\epsilon$.}\label{fig:massless_lowes_branches}
\end{figure}

\begin{figure}
    \centering
    \includegraphics[width=0.75\textwidth]{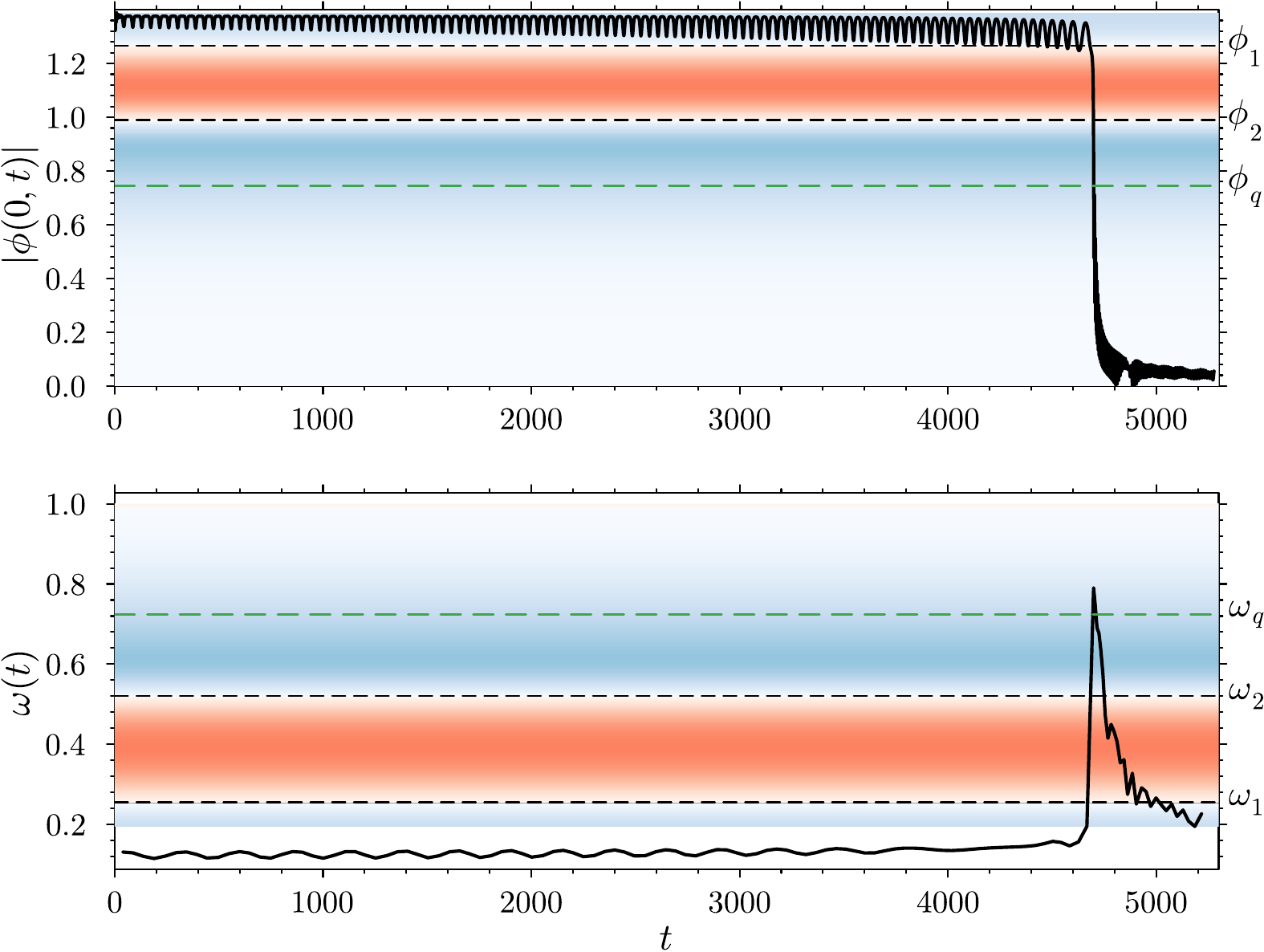}
    \caption{Q-ball decay in the massless rational model for $\omega=0.2$, $\epsilon=0.1$ -- different stages.
    }\label{fig:decay_massles_eps01 }
\end{figure}

\begin{figure}
    \centering
    \includegraphics[width=0.75\textwidth]{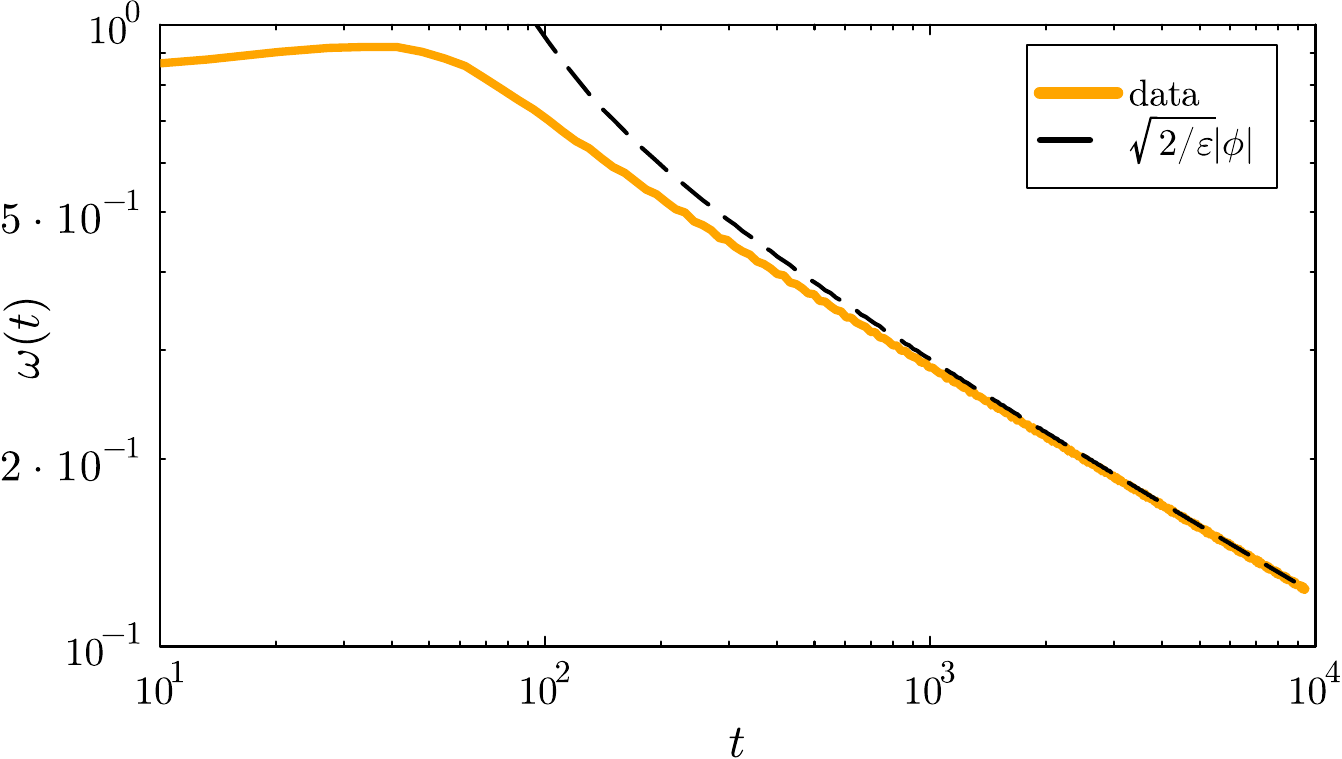}
    \caption{Last stage of the decay of a Q-ball in the massless model for $\epsilon=0.02$ and initial condition corresponding to a QB with $\omega=0.8$.
    }\label{fig:effective_mass_decay}
\end{figure}

\begin{figure}
    \centering
    \includegraphics[width=0.75\textwidth]{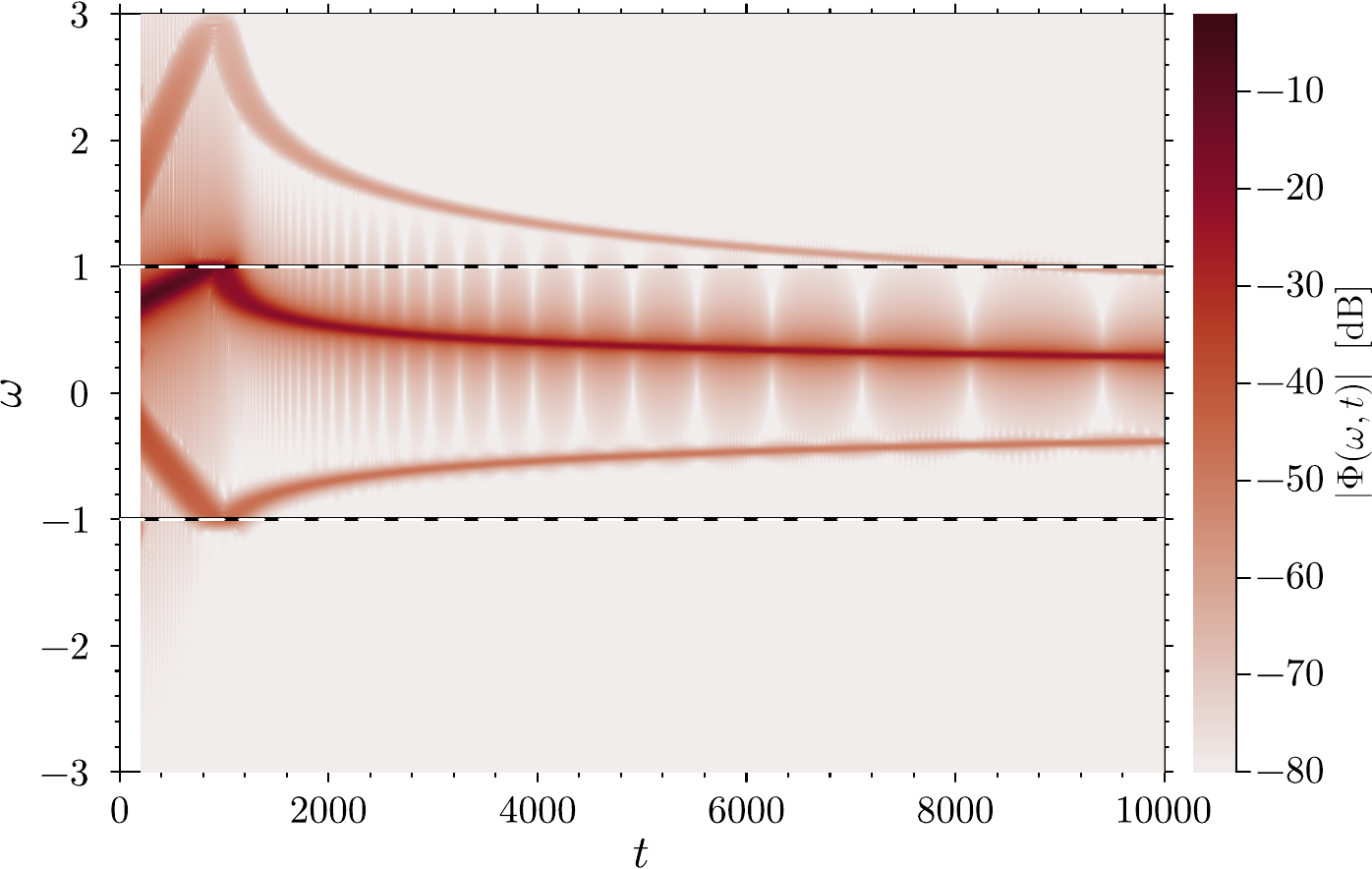}
    \caption{Spectrogram of an example evolution of the QQB in the massless model for $\epsilon=0.0025$ and initial conditions corresponding to a QB with $\omega=0.8$.
    }\label{fig:massless_spectrogram}
\end{figure}

In Figs.~\ref{fig:decay_massles_eps01 }-\ref{fig:massless_spectrogram} we present the results of the analysis of the time evolution of the Q-balls in the massless model \re{lag},\re{deform} for initial conditions corresponded to stationary solutions of the standard QB for $\epsilon=0$.
First, we observe that for large amplitude Q-balls there is a certain similarity between the radiative decay of the QQBs in a box, and the time evolution of the configuration in the massless
deformed model, the QQB gradually lose its energy mainly via radiation through the fundamental frequency.

Indeed, comparing the Figs.~ \ref{fig:decay02_domain} and \ref{fig:decay_massles_eps01 } we can see that in both cases a rapid transition happens 
after a relative long period of slow decrease of the amplitude of the radiating configurations.

In the last stage of the Q-balls in the massless model decay is very different from the last stage of the decay of a boxed Q-ball. In the massless model the frequency decreases along with the amplitude. This can be qualitatively understood through a notion of the effective mass threshold. In the limit $\phi\to 0$ the equation of motion can be expanded in the Taylor series. Keeping the lowest non-derivative term it becomes:
\begin{equation}
    \phi_{tt}-\phi_{xx}+\frac{2|\phi^2|}{\epsilon}\phi=0\,.
\end{equation}
For the field modulus of which varies slowly, the term 
\begin{equation}
    m_{\rm eff} = \sqrt{\frac{2}{\epsilon}}|\phi|
\end{equation}
can be regarded as a mean field effective mass parameter. Any small perturbation or perturbations changing only the phase of the field would behave as a field with this mass. This value can be treated as the upper limit of a temporary frequency for the quasi-QB. And indeed, numerical simulations show that as the QQB decays its frequency tends to this value from below, see Figure \ref{fig:effective_mass_decay}. The frequency also obeys nice scaling properties (indicated as a straight line on the log scale in Figure \ref{fig:effective_mass_decay})
\begin{equation}
    \omega(t)\sim t^{-\alpha}\,,\qquad |\phi(t)|\sim t^{-\alpha}\,,\qquad\alpha=0.36686\pm0.0006 
\end{equation}

\begin{figure}
    \centering
    \includegraphics[width=1\textwidth]{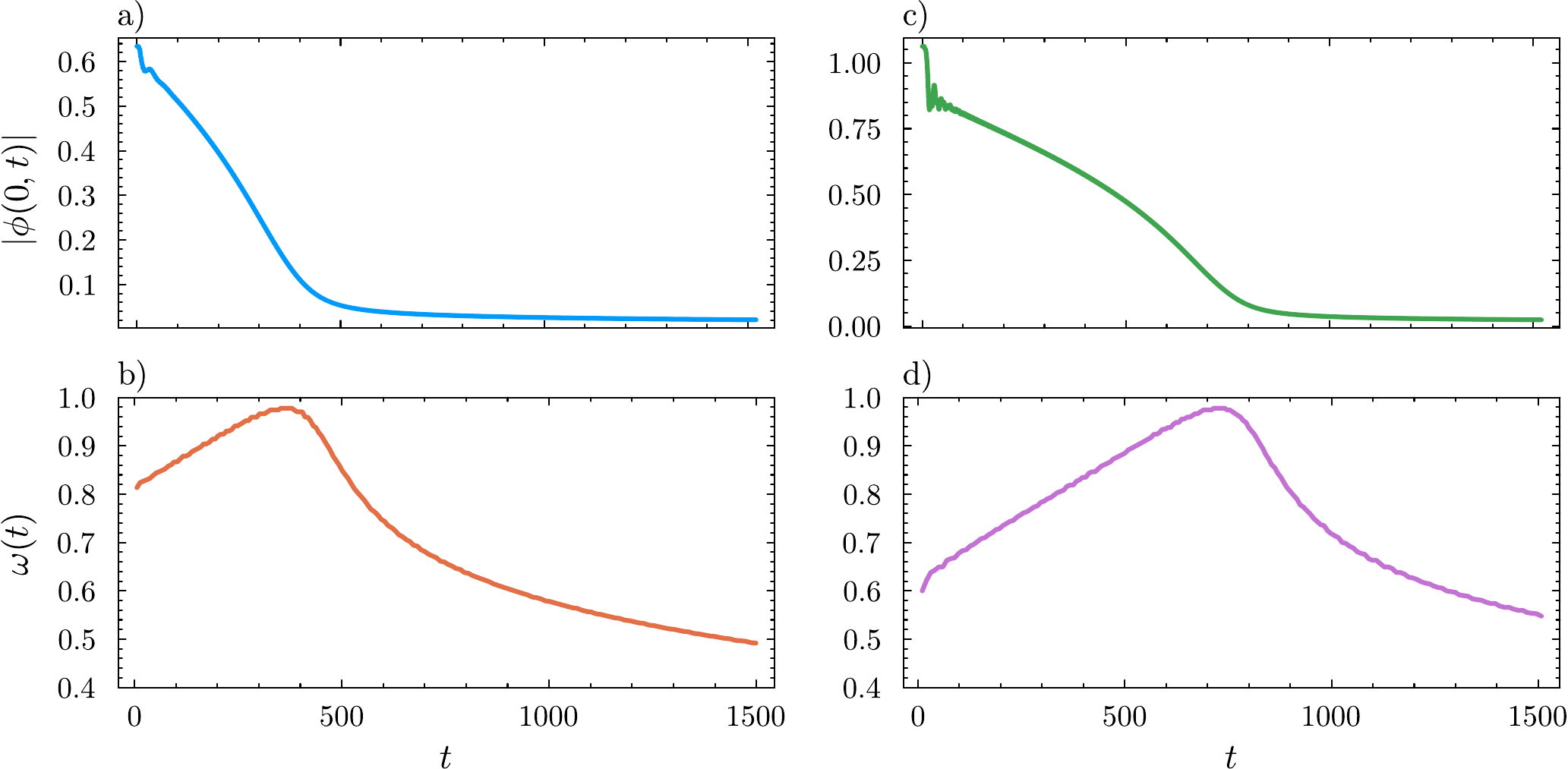}
    \caption{QQB decay in the rational model for $\epsilon=0.003$ and $\omega=0.6$ (a,b) and  $\omega=0.45$ (c,d).}\label{fig:decay06_massless}
\end{figure}

Our last Figure, \ref{fig:massless_spectrogram} shows the time-frequency spectrogram of the field at the centre for a QQB for a small parameter $\epsilon=0.0025$. Apart from the main frequency of the QQB there are two symmetrically placed peaks corresponding to a QNM of a standard QB indicating that initial conditions gave rise to a perturbed QQB. What is interesting is that the QNM remains visible even in the later stage of the evolution, when the fundamental frequency of the QQB decreas. Note that at this stage the QQB is very different from the standard QB and is itself a \textit{nonlinear} resonance. 

\section{Conclusions}

In this paper we have discussed long-time evolution of radiating  quasi-Q-balls.
Unlike usual Q-balls, which are stationary non-topological solitons in a theory of complex scalar field, the frequency of the QQBs is complex-valued. In that sense the QQBs  correspond to  slowly decaying nonlinear resonances.

Two models are considered; (i) The model without a mass gap with a deformed potential and (ii) Quasi-Q-ball in a box with outgoing boundary conditions. 
Both models provide a similar pattern of radiative decay of the large amplitude QQBs, however, differences appear at the last steps of decay. In both models initially the QQBs evolve adiabatically thorough nearly stationary states, similar to the original Q-balls. Due to radiation their frequencies increase and amplitudes decrease. QQBs in a box decay into linear QNMs, so at the late stage of the decay, their frequencies and profile widths become constant and their amplitudes decay exponentially with time. On the other hand, QQBs in the rational massless model in the late stage of their evolution the frequencies and amplitudes decrease but the widths increase, obeying certain scaling power laws. Therefore the non-linearity is crucial even in the late stage of the decay in the massless model.

Large amplitude, small amfrequency QQBs, corresponding to the lower stable branch, have particularly interesting evolution. Initially they lose energy due to the radiation until they reach the end of the stable branch. Then a sudden transition occurs and the QQBs land on the second stable branch, with a very similar charge to the one from before the transition. Energy of the QQBs is less, and its excess is allocated for example in the long lived QNMs of the QQBs. Surprisingly, the QNMs are still visible in the late stage of the evolution of the QQBs in the massless model.

We have also identified other branches of solutions satisfying outgoing boundary conditions. They have much shorter life-times and their role in the dynamics, at this point, is unclear. 

We expect that many of the features described in the present paper would remain in higher dimensions. However, the key difference between 1+1d and 3+1d Q-balls is the existence of an upper limit for the frequency  $\omega_{\textrm{max}}$ below the mass threshold. Therefore we expect that for some choice of parameters, the QQB would vanish quickly after reaching this frequency. This stage should be similar to the last stage of the massless QQB in 1+1d but with different scaling properties.

The exploration performed in this work has really just scratched the surface of a rich structure
of long-lime evolution and radiative decay of Q-balls and their relations with oscillons, which certainly merits to be analysed in greater depth.

\section*{Acknowledgements}
Y.S. gratefully acknowledges the support by FAPESP, 
project No 2024/01704-6 and thanks the Instituto de F\'{i}sica de S\~{a}o Carlos; IFSC for kind hospitality.
TR was supported by the Priority Research Area under the program Excellence Initiative—Research University at
the Jagiellonian University in Kraków.

\bibliographystyle{JHEP}
\bibliography{refs}
\end{document}